\documentstyle[pra,aps]{revtex}
\begin{document}
\preprint{}
\draft
%
\title{Dynamical suppression of decoherence \\ 
in two-state quantum systems}
\author{Lorenza Viola\thanks{Electronic address: vlorenza@mit.edu} 
and 
Seth Lloyd\thanks{Electronic address: slloyd@mit.edu} }
\address{ d'Arbeloff Laboratory for Information Systems and Technology, \\
Department of Mechanical Engineering, Massachusetts Institute of 
Technology, \\ Cambridge, Massachusetts 02139 } 
\date{\today}
\maketitle
%
%
\begin{abstract}
The dynamics of a decohering two-level system driven by a suitable control
Hamiltonian is studied. The control procedure is implemented as a sequence of
radiofrequency pulses that repetitively flip the state of the system,
a technique that can be termed quantum ``bang-bang'' control after
its classical analog. Decoherence introduced by the system's interaction 
with a quantum environment is shown to be washed out completely in the 
limit of continuous flipping and greatly suppressed provided the 
interval between the pulses is made comparable to the correlation 
time of the environment. The model suggests a strategy to 
fight against decoherence that complements existing quantum 
error-correction techniques.
\end{abstract}
%
%
%
%
\section{Introduction}

The design of strategies able to protect the evolution of a quantum system
against the irreversible corruption due to environmental noise represents a 
challenging conceptual issue.  In particular, since maintaining quantum 
coherence is a crucial requirement for exploiting the novel possibilities
opened up by quantum parallelism, practical implementations 
of quantum computation and communication proposals require methods to 
effectively resist the action of quantum decoherence and dissipation 
\cite{review,unruh,palma}. Roughly speaking, two classes of procedures are
available to overcome the decoherence problem: either passive 
stabilization or active manipulation of the quantum state. The first 
kind of solutions, recently formalized as {\sl Error-Avoiding Codes} 
\cite{rasetti}, relies on the existence of a subspace of states that, owing 
to special symmetry properties, are dynamically decoupled from the 
environment. The second approach, pionereed in \cite{shor} and closer in 
spirit to quantum control theory \cite{butko,seth1}, embraces today a 
variety of sophisticated schemes known as {\sl Error-Correcting Codes} 
\cite{steane,calderbank,laflamme,bennett,gottesman,cirac,divincenzo,schumacher,vaidman,luming,slotine}. 
Basically, loss of information is corrected by monitoring the system and 
conditionally carrying on suitable feedback operations.

In this work we investigate a third strategy for reducing noise and
decoherence. This strategy, which can be termed quantum ``bang-bang'' control
after its classical analog \cite{bang},  works by averaging out the unwanted 
effects of the environmental interaction through the application of suitable 
open-loop control techniques on the system.
The basic idea is that open-system properties, specifically decoherence,
may be modified if a time-varying control field acts on the dynamics of the
system over time scales that are comparable to the memory time of the 
environment. In particular, we work out an exact model for a two-state quantum 
system ({\sl qubit}) coupled to a thermal bath of harmonic oscillators, 
where decoherence is dynamically suppressed through repeated effective 
time-reversal operations on the combined system $+$ bath. 
Although the phenomenon is mathematically reminiscent of the quantum Zeno 
effect \cite{zeno}, the essential physical idea comes from refocusing 
techniques in nuclear magnetic resonance spectroscopy (NMR) \cite{slichter}.
Since the discovery of spin echoes in 1950 \cite{hahn}, 
clever pulse methods have been developed in NMR to eliminate much of the 
dephasing arising from variations in the local magnetic field acting on each
spin.  Because the latter effect can be thought in terms of an interaction with 
some classical environment, it is not obvious {\sl a priori} whether similar 
techniques work in the presence of a quantum mechanical environment and 
purely nonclassical effects like entanglement. Our result answers 
this question in the affirmative, and points out the role of the reservoir 
correlation time as a further parameter to be engineered in the struggle for
preserving quantum coherence.

The plan of the paper is the following. In Section II, a general model of 
a two-state system interacting with a thermal environment is reviewed 
and its decoherence properties in the absence of control recalled. In 
Section III, the evolution under the action of a sequence of perturbative  
kicks is analyzed. Complete quenching of decoherence is established as a 
limiting situation. In Section IV, the conditions for an effective 
decoherence reduction are clarified with reference to a variety of possible 
environmental configurations and the method is compared to different 
quantum error-correction techniques. We close by discussing possibilities 
for future work.  

\section{Single-qubit dephasing mechanism}

Our goal is to investigate how decoherence properties of an open quantum 
system may be modified through the application of an external controllable
interaction.  Decoherence is a process whereby quantum systems
lose their ability to exhibit coherent behavior such as interference
\cite{gellmann,zurek,griffiths,omnes}.   
We start by introducing a model that allows investigation of the 
problem in its simplest nontrivial configuration. The physical system we are
interested in is a single two-state quantum system, representing the 
elementary memory cell of quantum information (qubit). 
Although not strictly necessary, it will be convenient to think of the 
physical qubit as realized by a spin-$1/2$ system, which will provide us 
with direct reference to the well established language of Nuclear Magnetic 
Resonance \cite{slichter} and the rapidly growing field of NMR quantum 
computation \cite{nmr,nmr1}. 
Decoherence arises due to the coupling to a quantized environment, here 
schematized as a continuum of harmonic modes. We assume that the 
dynamics of the overall qubit $+$ bath is ruled by the following 
Hamiltonian:
\begin{equation}
H_0 = H_S + H_B +H_{SB} = \hbar \omega_0 \, {\sigma_z \over 2}  + 
\sum_k \, \hbar \omega_k b_k^\dagger b_k + 
\sum_k \, \hbar \sigma_z (g_k b_k^\dagger + g_k^\ast b_k) \;,    
\label{hamiltonian}
\end{equation}
where the first and second contribution $H_S$ and $H_B$ describe, 
respectively,
the free evolution of the qubit and the environment, and the third term 
$H_{SB}$ describes a bilinear interaction between the two. 
$\sigma_z$ is the standard diagonal Pauli matrix, with qubit basis states 
denoted as $|i \rangle$, $i=0,1$, while $b_k^\dagger,b_k$ are bosonic
operators for the $k^{\text{\small th}}$ field mode, characterized by a 
generally complex coupling parameter $g_k$.
In the Schr\"{o}dinger picture, the state of the combined system $(S+B)$ is
represented by a density operator $\rho_{tot}(t)$ and the reduced qubit 
dynamics is thereupon recovered from a partial trace over the environment 
degrees of freedom:
\begin{equation}
\rho_S(t)=\sum_{i,j=0,1} \, \rho_{ij}(t) \,|i \rangle \langle j| = 
\mbox{Tr}_B \{ \rho_{tot}(t) \} \;.
\label{trace}
\end{equation}
Hamiltonian (\ref{hamiltonian}), which corresponds to a special case of the 
so-called spin-boson problem \cite{leggett}, has been used by many authors to 
model decoherence in quantum computers \cite{unruh,palma,rasetti}. 
In particular,
we adhere closely to the notations of \cite{palma}. The basic fact about the 
dynamics induced by (\ref{hamiltonian}) is that, since $[\sigma_z, H_0]=0$, the
interaction with the environment has the two memory states 
$|0\rangle, |1\rangle$ 
as eigenstates. In other words, the model describes a purely decohering 
mechanism, where no energy exchange between qubit and bath is present. In NMR 
terminology, this implies that no $T_1$-type of decay takes place 
\cite{slichter}.
Equivalently, in terms of errors, only phase errors are introduced. However, 
neglecting the effects associated to quantum dissipation is
justified, in a sense, by two related reasons: energy exchange processes not 
only produces amplitude errors which need to be corrected even in the 
classical computation, but they typically involve time scales much longer than 
decoherence mechanisms. In addition, being exactly soluble, the model 
(\ref{hamiltonian}) has the advantage of allowing a clear picture of the 
decoherence properties in the absence of control. To this end, since spin 
populations are not affected by the 
environment, the relevant quantity is the qubit coherence 
$\rho_{01}(t)$ (of course, $\rho_{10}(t)=\rho_{01}^\ast(t)$).  

It will be convenient to move to the interaction picture associated to the 
free dynamics $(H_S+H_B)$, corresponding to the transformed state vector
\begin{equation}
\tilde{\rho}_{tot}(t)=\mbox{e}^{i (H_S+H_B)t/\hbar} \, \rho_{tot}(t) \, 
\mbox{e}^{-i (H_S+H_B)t/\hbar}
\label{interaction}
\end{equation}
and to the effective Hamiltonian
\begin{equation}
\tilde{H}(t)=\tilde{H}_{SB}(t)= \hbar\sigma_z \sum_k \,
\Big( g_k b_k^\dagger \mbox{e}^{i \omega_k t} + 
  g_k^\ast b_k \mbox{e}^{-i \omega_k t} \Big) \;. 
\label{effective}
\end{equation}
Time evolution is determined by the time-ordered unitary operator
\begin{equation}
\tilde{U}_{tot}(t_0,t)=T\exp\bigg\{ -{i \over \hbar} \int_{t_0}^t 
ds \, \tilde{H}(s) \bigg\} \;, 
\label{timeprod}
\end{equation}
which can be evaluated exactly and can be written, up to a global $c$-number 
phase factor, in the following form:
\begin{equation}
\tilde{U}_{tot}(t_0,t)= \exp\bigg\{  {\sigma_z \over 2} 
\sum_k \Big( b_k^\dagger \mbox{e}^{i \omega_k t_0} \xi_k(t-t_0) -
         b_k \mbox{e}^{-i \omega_k t_0} \xi_k^\ast(t-t_0) \Big) \bigg\} \;, 
\label{timeop1}
\end{equation}
where the following complex function has been introduced:
\begin{equation}
\xi_k(\Delta t) = {2 g_k \over \omega_k} \,  
\Big( 1- \mbox{e}^{i \omega_k \Delta t} \Big) \;.
\label{csi}
\end{equation}
Note that the separate dependence of (\ref{timeop1}) on both initial time 
$t_0$ and evolution interval $(t-t_0)$ is consistent with the time-reversal 
property $\tilde{U}_{tot}(t,t_0)\tilde{U}_{tot}(t_0,t)=\openone$. A nice 
discussion on the 
entanglement generated by $\tilde{U}_{tot}(t_0,t)$ between qubit and field 
states is given in \cite{palma}. We are interested in calculating 
\begin{equation}
\tilde{\rho}_{01}(t)=\langle 0|\, \mbox{Tr}_B \{ \, 
\tilde{U}_{tot}(t_0,t) \tilde{\rho}_{tot}(t_0) \tilde{U}_{tot}^\dagger(t_0,t)
\, \} \, |1 \rangle \;. 
\label{rho01}
\end{equation}
This can be done without approximations under two standard assumptions:  
\par\noindent
$(i)\;$ qubit and environment are initially uncorrelated, i.e.
\begin{equation}
\rho_{tot}(t_0)= \rho_S(t_0) \otimes \rho_B(t_0) \;;
\label{factorized}
\end{equation}
\noindent 
$(ii)$ the environment is initially in thermal equilibrium at temperature 
$T$, i.e. 
\begin{equation}
\rho_B(t_0)= \prod_k \, \rho_{B,k}(T) = \prod_k \,
\Big( 1- \mbox{e}^{\beta \hbar \omega_k} \Big) \, 
\mbox{e}^{-\beta \hbar \omega_k b_k^\dagger b_k } \;. 
\label{thermal}
\end{equation}
In Eq. (\ref{thermal}), $\beta= 1/k_B\, T$, being $k_B$ the Boltzmann constant.
For simplicity, we choose henceforth units such that $\hbar =k_B =1$.
Conditions (\ref{factorized})-(\ref{thermal}) are easily translated to 
interaction picture and, inserting (\ref{timeop1}) into (\ref{rho01}), the
problem is reduced to a single-mode trace:
\begin{equation} 
\tilde{\rho}_{01}(t) = \tilde{\rho}_{01}(t_0) \cdot \prod_k \,
\mbox{Tr}_k \Big\{\, \rho_{B,k}(T) 
\, {\cal D} ( \mbox{e}^{i \omega_k t_0} \xi_k(t-t_0) ) \, \Big\}
= \tilde{\rho}_{01}(t_0) \cdot \mbox{e}^{ - \Gamma_0(t_0,t) }
\;, \label{singletrace}
\end{equation}
where, in the first equality, the harmonic displacement operator 
${\cal D}(\xi_k)$ is given by 
\begin{equation}
{\cal D}(\xi_k) = \mbox{e}^{\, b_k^\dagger \xi_k - b_k \xi_k^\ast} \;,  
\label{displace}
\end{equation}
and the second equality defines the time-dependent function 
$\Gamma_0(t_0,t)$. The final step is to recognize that, for each mode, the 
quantity in curly brackets in (\ref{singletrace}) is nothing but the 
symmetric order generating function for a thermal harmonic oscillator
\cite{hillery}. Thus, the explicit expression for $\Gamma_0(t_0,t)$ is
\begin{equation}
\Gamma_0(t_0,t)=\Gamma_0(t-t_0) = \sum_k \, { |\xi_k(t-t_0)|^2 \over 2}
\coth \bigg({\omega_k \over 2T} \bigg) \;. 
\label{gamma0}
\end{equation}
Being $\Gamma_0$ a real function, it correponds to pure damping in 
(\ref{singletrace}). Accordingly, this function characterizes completely 
the dynamics of the decoherence process destroying the qubit phase 
information. Of course, the complete evolution in the original 
Schr\"odinger picture includes oscillation at the natural frequency 
$\omega_0$,
\begin{equation}
\rho_{01}(t)=\mbox{e}^{i \omega_0 (t-t_0) - \Gamma_0(t-t_0)} \, 
\rho_{01}(t_0) \;, 
\label{lab}
\end{equation}
while, using (\ref{timeop1}), one can check that 
$\rho_{ii}(t)=\rho_{ii}(t_0)$, $i=0,1$.

Deeper insight into the time dependence of the decoherence process is 
gained if the continuum limit is made explicit in (\ref{gamma0}). By 
substituting $|\xi_k(t-t_0)|^2$ through (\ref{csi}), we get
\begin{equation}
\Gamma_0 (t-t_0) = \int_0^\infty d\omega \, 
\bigg[ \sum_k \delta(\omega-\omega_k) |g_k|^2 \bigg]\,
4 \coth\bigg( {\omega \over 2T} \bigg) 
{1-\cos \omega(t-t_0) \over \omega^2} \;.
\label{spectral}
\end{equation}
The quantity in square brackets is known as the spectral density $I(\omega)$ 
of the bath. It turns out that, once the initial state is specified, complete
information about the effect of the environment is encapsulated in this 
single function. As a general feature, the spectral density is characterized 
by a certain ultraviolet cut-off frequency $\omega_c$ such that 
$I(\omega) \rightarrow 0$ for $\omega > \omega_c$. Although the specific 
value of $\omega_c$ depends on a natural cut-off frequency varying 
from system to system, the existence of a finite $\omega_c$ is always
demanded on physical grounds. Indeed, assuming that the environment does not
have a high frequency cut-off means, generally, that energy can be dissipated
instantaneously \cite{paz}. If, for instance, decoherence arises from the 
coupling to a phonon field, the natural cut-off frequency can be identified 
with the Debye frequency. In general, $\tau_c \sim \omega_c^{-1}$
sets the fastest time scale (or the memory time) of the environment. When a 
specific choice will be useful, we will assume a spectral density with the 
following functional form:
\begin{equation}
I(\omega) = {\alpha \over 4} \, \omega^n \, \mbox{e}^{-\omega/\omega_c} 
\;,  \label{density}
\end{equation}
the parameter $\alpha > 0$ measuring (in suitable units) the strength of the 
system-bath coupling  and the index $n>0$ classifying different environment
behaviors \cite{palma,leggett,paz}.

In addition to $\tau_c$, another time scale $\tau_{\beta} \sim T^{-1}$,
associated with the temperature of the bath, is expected to play a major role
in the evolution of the qubit coherence. This is manifest if Eq. 
(\ref{spectral}) is written in the equivalent form
\begin{equation}
\Gamma_0 (t-t_0) = 4\, \int_0^\infty d\omega \, I(\omega)\, 
\Big( 2\, \overline{n}(\omega,T) +1 \Big)\,  
{1-\cos \omega(t-t_0) \over \omega^2} \;,
\label{spectral2}
\end{equation}   
where $\overline{n}(\omega,T)=\exp(-\omega/2T) 
\mbox{cosech}(\omega/2T)$ is the 
average number of field excitations at temperature $T$. In this way, effects
due to the thermal noise are formally separated from the ones due to purely
vacuum fluctuations. Because of the different frequency composition, the 
two kinds of fluctuations dominate on different time scales, the 
relative importance of vacuum to thermal contributions being determined 
by $\tau_{\beta}$. This multiplicity of time scales is one of the factors 
that makes the decoherence process quite complicated. It should be important
to keep in mind that there is {\sl no} generic decohering behavior, and in 
particular that the qubit dynamics depends crucially on both temperature and
the details of the spectral function $I(\omega)$. We will comment further on
this point later.

\section{Pulsed evolution of quantum coherence}

Let us now define a procedure aimed at modifying the decoherence properties
discussed so far. We choose to implement it as a suitable perturbation 
acting on some observables $\{ {\cal O}_i\}$ of system $S$. This is 
obtained by adding to (\ref{hamiltonian}) a time-dependent term 
$\sum_i \, \gamma_i(t) {\cal O}_i$, where the input functions 
$\{\gamma_i(t)\}$ are assumed to be programmable at will, e.g. 
a given schedule of time-varying magnetic fields in the case of a spin-qubit.
In control terminology, this realizes a so-called open-loop configuration
\cite{seth1}. Closed-loop (or feedback) configurations have 
been also proposed in recent years to manipulate decoherence in some quantum
optical systems \cite{milburn}. There are many possible choices for the 
control Hamiltonian. Leaving aside the problem of controllability of the 
system (\ref{hamiltonian}) at the abstract level, we make here a pragmatic 
choice, partly suggested by semiclassical considerations. 

Suppose that the perturbation we add is able to induce spin-flip transitions.
By inspection of the spin-bath interaction Hamiltonian $H_{SB}$, opposite 
contributions arise when the spin belongs to the down or up eigenstate. Since 
a relative minus sign will be present during time evolution, the effect of the 
$H_{SB}$ coupling will eventually average out provided the spin is flipped 
rapidly enough. There are two reasons why a mechanism of this kind is expected
to be possible. First, as we already mentioned, similar methods are 
routinely used in NMR experiments to get rid of the effects of some unwanted 
interactions \cite{slichter}. For instance, in the so-called spin-flip 
narrowing, line broadening resulting from magnetic dipolar coupling is reduced
by repetitively flipping the spins among distinct energy configurations. 
The major difference here is that the undesired decohering coupling is 
contributed by the infinitely-many quantized degrees of freedom of the 
heat bath. The second reason is related to the finite response time $\tau_c$
of the environment itself. In general, if perturbations act on the system
more rapidly than this fastest time scale present in the bath, we expect that
memory effects would be relevant, and that this non-Markovian 
dynamics would eventually lead to modified decoherence properties. 

Having anticipated the intuitive idea, we start by writing the total 
Hamiltonian as
\begin{equation}
H(t)=H_0 + H_{rf}(\omega_0,t) \;, 
\label{ham2}
\end{equation}
where $H_{rf}(\omega_0,t)$ represents a monochromatic alternating magnetic 
field applied at resonance. At this level, the problem can
be related to variants of the complete spin-boson model, none of them 
is exactly solvable \cite{leggett}. The main simplification we 
introduce is to substitute the continuous-mode operation, where the actions
of the bath and the controlling field are necessarily simultaneous, with a
pulsed-mode operation analogous to classical ``bang-bang'' control. 
Then, under the working assumption that the typical 
decoherence time of the system and the duration of pulses define two widely 
different time scales, we solve the problem by putting
\par\noindent $(i)\;$ $H_{SB}=0$ within each pulse; 
\par\noindent $(ii)$ $H_{rf}=0$ between successive pulses. 
\par\noindent As usual, we invoke the rotating wave approximation and only 
look at the co-rotating component of the rf-field.   
The radiofrequency perturbation is assumed of the following form:
\begin{equation}
H_{rf}(\omega_0,t)= \sum_{n=1}^{n_P}\, V^{(n)}(t) \Big[
\cos \Big( \omega_0(t-t^{(n)}_P)\Big) \sigma_x + 
\sin  \Big( \omega_0(t-t^{(n)}_P)\Big) \sigma_y \Big] \;, 
\label{rf}
\end{equation}
with $t^{(n)}_P = t_0 + n \Delta t$, $n=1,\ldots,n_P$, and 
\begin{equation}
V^{(n)}(t) = \left\{ \begin{array}{lc}
           V \hspace{3mm} & t^{(n)}_P \leq t \leq t^{(n)}_P + \tau_P \;, \\
           0 \hspace{3mm} & \text{elsewhere} \;. 
\end{array}  \right.
\label{V}
\end{equation}
Eqs. (\ref{rf})-(\ref{V}) schematize a sequence of $n_P$ identical pulses,
each of duration $\tau_P$, applied at instants $t=t^{(n)}_P$. The separation
$\Delta t$ between pulses is assumed to be an input of the model. The 
amplitude of the field is equal to $V$ during each pulse and will be further
specified below together with $\tau_P$. In order to depict the evolution
associated to a given pulse sequence, it is convenient to think the latter as
formed by successive elementary cycles of spin-flips, a complete cycle being
able to return the spin back to the starting configuration. We begin by 
analyzing the time evolution during the first spin cycle. 
  
\begin{center}
{\bf A. The elementary spin-flip cycle}
\end{center}

As in Sec. II, we exploit the interaction representation 
(\ref{interaction}). Some instructive steps of an alternative derivation based 
on Heisenberg formalism are sketched in the Appendix. The interaction picture 
transformed Hamiltonian is now 
\begin{equation}
\tilde{H}(t)=\tilde{H}_{SB}(t) + \tilde{H}_{rf}(t) \;, 
\label{effective2}
\end{equation}
where $\tilde{H}_{SB}(t)$ is given in (\ref{effective}), and 
$\tilde{H}_{rf}(t)$
is evaluated by using (\ref{rf}) and the properties of Pauli matrices. We get
\begin{equation}
\tilde{H}_{rf}(t)= \sum_{n=1}^{n_P} \, V^{(n)}(t) \,
\mbox{e}^{i \omega_0 {\sigma_z \over 2} t^{(n)}_P } \sigma_x
\mbox{e}^{-i \omega_0 {\sigma_z \over 2} t^{(n)}_P } \;. 
\label{rftilde}
\end{equation}
According to (\ref{rftilde}), the time dependence due to the rotating field is
completely removed within each pulse. In fact, the interaction representation 
(\ref{interaction}) on spin variables is identical, at resonance, with the 
description in the rotating frame associated to (\ref{rf}) \cite{slichter}.
The counter-rotating term that is omitted within RWA is seen to be  
negligible at resonance. For the first spin 
cycle, $n_P=2$ and we have the following sequence: evolution under 
$\tilde{H}_{SB}(t)$ during $t_0 \leq t \leq t^{(1)}_P$; pulse $P_1$ at time
$t^{(1)}_P$; evolution under $\tilde{H}_{SB}(t)$ during 
$t_P^{(1)}+\tau_P \leq t \leq t^{(2)}_P$; pulse $P_2$ at time $t^{(2)}_P$. 
After a total time $t_1=t_0 + 2 \Delta t+ 2 \tau_P$, the first cycle is 
complete. 
In terms of evolution operators, we have:
\begin{equation}
\tilde{U}_P(t_0,t_1)= \tilde{U}_{P_2} \tilde{U}_{P_1} \,\Big[
\tilde{U}_{P_1}^{-1} \tilde{U}_{tot}(t^{(1)}_P + \tau_P, t^{(2)}_P) 
\tilde{U}_{P_1} \Big] \Big[ \tilde{U}_{tot}(t_0,t^{(1)}_P) \Big ] \;. 
\label{timeop2}
\end{equation} 
We can read the evolutions in the absence of rf-field directly from 
(\ref{timeop1}), for instance 
\begin{equation}
\tilde{U}_{tot}(t_P^{(1)}+\tau_P,t^{(2)}_P)= \exp\bigg\{  {\sigma_z \over 2} 
\sum_k \Big( b_k^\dagger \mbox{e}^{i \omega_k (t_0+\Delta t+ \tau_P)} 
\xi_k(\Delta t) - b_k \mbox{e}^{-i \omega_k (t_0+\Delta t + \tau_P)} 
\xi_k^\ast(\Delta t) \Big) \bigg\} \;.
\end{equation}
Concerning the evolution operator associated to the generic 
$j^{\text{\small th}}$ pulse, this is found by exponentiating (\ref{rftilde}) 
($n=j$):
\begin{equation}
\tilde{U}_{P_j}=\exp\Big\{ -i \,V \tau_P \,
\mbox{e}^{i \omega_0 {\sigma_z \over 2} t^{(j)}_P } \sigma_x
\mbox{e}^{-i \omega_0 {\sigma_z \over 2} t^{(j)}_P } \Big\} =
\mbox{e}^{i \omega_0 {\sigma_z \over 2} t^{(j)}_P }
\mbox{e}^{-i V \tau_P \sigma_x} 
\mbox{e}^{-i \omega_0 {\sigma_z \over 2} t^{(j)}_P } \;. 
\label{timep}
\end{equation}
In order to proceed, we have to say more about pulses. We require
\begin{equation}
\{ \tilde{U}_{P_1}, \sigma_z \} =0 \;, \hspace{3cm} 
[ \tilde{U}_{P_2} \tilde{U}_{P_1}, \sigma_z] =0 \;, 
\label{pulses}
\end{equation}
$\{\,,\,\}$ and $[\,,\,]$ denoting anticommutator and commutator respectively.
These conditions imply, as expected from the intuitive explanation, that 
$P_1,P_2$ are $\pi$-pulses, satisfying $2\,V\tau_P= \pm \, \pi$. 
To simplify things, we imagine that $\tilde{H}_{rf}(t)$ is large enough to 
produce (almost) instantaneous spin-flips. Accordingly, $\tau_P \rightarrow 0$ 
henceforth and we have to deal with ideal ``kicks'' of infinite power. Putting
things together, we find
\begin{equation}
\tilde{U}_{P_2}\tilde{U}_{P_1} = - \, \mbox{e}^{i \omega_0 {\sigma_z \over 2}
(t_1-t_0) } \;, 
\end{equation}
\begin{equation}
\tilde{U}_{P_1}^{-1} \tilde{U}_{tot}(t^{(1)}_P, t^{(2)}_P) 
\tilde{U}_{P_1} = 
\exp\bigg\{ - {\sigma_z \over 2} 
\sum_k \Big( b_k^\dagger \mbox{e}^{i \omega_k (t_0+\Delta t)} 
\xi_k(\Delta t) - b_k \mbox{e}^{-i \omega_k (t_0+\Delta t)} 
\xi_k^\ast(\Delta t) \Big) \bigg\} \;,
\label{transformed}
\end{equation}
and we are therefore ready to write down the final result for the cycle 
evolution operator:
\begin{equation}
\tilde{U}_P(t_0,t_1) = 
\exp\bigg\{ i  \omega_0 \, {\sigma_z \over 2} (t_1-t_0) + 
 {\sigma_z \over 2} 
\sum_k \Big( b_k^\dagger \mbox{e}^{i \omega_k t_0} 
\eta_k(\Delta t) - b_k \mbox{e}^{-i \omega_k t_0} 
\eta_k^\ast(\Delta t) \Big) \bigg\} \;,
\label{pulsed}
\end{equation}
where, as before, $c$-number phase factors have been omitted and 
\begin{equation}
\eta_k(\Delta t) =\xi_k (\Delta t) 
\Big(\mbox{e}^{i \omega_k \Delta t} -1 \Big) = 
- {2 g_k \over \omega_k} \Big( 1-\mbox{e}^{i \omega_k \Delta t} \Big)^2 \;.
\label{eta}
\end{equation}
It is interesting to compare the evolution described by (\ref{pulsed}) with 
the one in the absence of pulses. By recalling (\ref{timeop1}), 
evaluated at time $t_1=t_0 + 2 \Delta t$, we report two differencies: a 
phase factor proportional to $\sigma_z$ and the duration of the cycle; a 
combination $\eta_k(\Delta t) \propto (\xi_k(\Delta t))^2$ in place of 
$\xi_k (2 \Delta t)$. The first difference corresponds to the fact that, due to
the pulses, the oscillation at the natural frequency $\omega_0$ is lost once 
the evolution is transferred back to Schr\"odinger picture (see (\ref{lab})).
The second difference, as it will be seen in a moment, is the signal that 
decoherence properties are modified.  
 
\begin{center}
{\bf B. Decoherence properties after a pulse sequence}
\end{center}

The next step is to generalize the description to an arbitrary number $N$ of
elementary spin-flip cycles, the $n^{\text{\small th}}$ cycle ending at time
\begin{equation}
t_n=t_0 + 2 n \Delta t \;, \hspace{3cm}n=1,\ldots,N\;, 
\label{endtime}
\end{equation}
and the number of $\pi$-pulses involved in the sequence being $n_P=2N$. This
is straight forward since Eq. (\ref{pulsed}) enables us to write down the 
evolution operator for the $n^{\text{\small th}}$ cycle:
\begin{equation}
\tilde{U}_P(t_{n-1},t_n) = 
\exp\bigg\{ i  \omega_0 \, {\sigma_z \over 2} 2\Delta t + 
 {\sigma_z \over 2} 
\sum_k \Big( b_k^\dagger \mbox{e}^{i \omega_k t_{n-1}} 
\eta_k(\Delta t) - b_k \mbox{e}^{-i \omega_k t_{n-1}} 
\eta_k^\ast(\Delta t) \Big) \bigg\} \;.
\label{pulsedn}
\end{equation}
The time development corresponding to $N$ cycles is then governed by the 
time-ordered finite product
\begin{equation}
\tilde{U}_P^{(N)}(t_0,\ldots,t_N) = \tilde{U}_P({t_{N-1},t_N}) \ldots  
\tilde{U}_P(t_1,t_2)
\tilde{U}_P(t_0,t_1) 
\;. \label{finitep}
\end{equation}
Note that, at variance with conventional spin-flip narrowing \cite{slichter}, 
$\tilde{U}^{(N)}_P \neq (\tilde{U}_P)^N$, the dependence on intermediate 
times $\{t_j\}$ in the sequence being introduced by the environment dynamics.
A closed formula for $\tilde{U}^{(N)}_P(\{t_j\})$ can still be found quickly 
since, by neglecting state-independent global phase factors that are irrelevant 
to density matrix propagation, we are allowed to treat the factors in 
(\ref{finitep}) as commuting operators. We get
\begin{equation}
\tilde{U}_P^{(N)}(t_0,\Delta t) = 
\exp\bigg\{ i  \omega_0 \, {\sigma_z \over 2} 2N \Delta t + 
 {\sigma_z \over 2} 
\sum_k \Big( b_k^\dagger \mbox{e}^{i \omega_k t_0} 
\eta_k(N,\Delta t) - b_k \mbox{e}^{-i \omega_k t_0} 
\eta_k^\ast(N,\Delta t) \Big) \bigg\} \;,
\label{sequence}
\end{equation}
where 
\begin{equation}
\eta_k(N,\Delta t) = \eta_k(\Delta t)\sum_{n=1}^N \,
\mbox{e}^{2i(n-1)\omega_k \Delta t}\;, 
\label{etak}
\end{equation}
and definition (\ref{endtime}) has been exploited. Of course, the results
of Sec. IIIA are recovered for $N=1$. A more interesting check can be done
by relating the evolution (\ref{sequence}) to the corresponding propagator in
the absence of pulses. The trick is to switch back the minus sign in the 
definition (\ref{eta}) of $\eta_k(\Delta t)$, that has been recognized as 
the key dynamical effect due to the pulsing procedure. Since
$\xi_k(\Delta t )(\mbox{e}^{i \omega_k \Delta t} +1)= \xi_k(2 \Delta t)$, we  
are left with 
\begin{equation}
\xi_k(2\Delta t )\sum_{n=1}^N \mbox{e}^{2i (n-1)\omega_k \Delta t} 
= \, \xi_k(2 N\Delta t) = \xi_k(t_N-t_0)
\end{equation}
in place of (\ref{etak}). Therefore, this procedure gives 
\begin{equation}
\tilde{U}_{tot}(t_0,t_N) = 
\exp\bigg\{  {\sigma_z \over 2} 
\sum_k \Big( b_k^\dagger \mbox{e}^{i \omega_k t_0} 
\xi_k(t_N-t_0) - b_k \mbox{e}^{-i \omega_k t_0} 
\xi_k^\ast(t_N-t_0) \Big) \bigg\} \;,
\end{equation} 
which does agree with the direct evaluation based on (\ref{timeprod}). It may
be worth some times to let the connection with unperturbed evolution be 
explicit, which is done by writing  
\begin{equation}
\eta_k(N,\Delta t)=\xi_k(2 N \Delta t) - 2 \,\xi_k (\Delta t) \sum_{n=1}^N
\mbox{e}^{2i (n-1)\omega_k \Delta t}  \;, 
\label{eta2}
\end{equation}
where the contribution due to the pulse sequence shows up in the form of a 
typical interference factor.

The decoherence properties corresponding to the pulsed qubit $+$ bath 
evolution (\ref{sequence}) are derived through the steps outlined in 
Sec. II for the unperturbed case. Incidentally, we still expect that spin
populations are unchanged after a sequence of $N$ complete spin cycles, while
qubit coherence at final time $t_N=t_0 + 2N \Delta t$ is given by 
\begin{equation}
\tilde{\rho}_{01}(t_N)=\langle 0|\, \mbox{Tr}_B \{ \, 
\tilde{U}_P^{(N)}(t_0,\Delta t) \tilde{\rho}_{tot}(t_0) 
\tilde{U}_P^{(N)\,\dagger} (t_0,\Delta t) \, \} \, |1 \rangle \;. 
\label{rho01p}
\end{equation}
The result is 
\begin{equation}
\tilde{\rho}_{01}(t_N)=\mbox{e}^{-i \omega_0 (t_N-t_0) - 
\Gamma_P(N,\Delta t)} \, \tilde{\rho}_{01}(t_0) \;, 
\label{lab2}
\end{equation}
where
\begin{equation}
\Gamma_P(N, \Delta t) = \sum_k \, { |\eta_k(N,\Delta t)|^2 \over 2} \,
\coth\bigg( {\omega_k \over 2T} \bigg) \;.
\label{gammap}
\end{equation}
Comparing with (\ref{gamma0}), the mathematical prescription for 
decoherence in the presence of pulses looks very simple: just use 
$\eta_k(N,\Delta t)$ instead of $\xi_k(t_N-t_0)$ for each mode of the bath. 
However, the final effect leading to (\ref{gammap}) is not easy to figure out 
and it is useful to compare first the decoherence due to a single mode with 
frequency $\omega$ with and without pulses respectively. Apart from identical 
time-independent factors, we have to consider
\begin{equation}
|\eta(N,\omega\Delta t)|^2 = 4 \, (1-\cos \omega\Delta t)^2 \, \Big[
N+ \sum_{n=0}^{N-1} \, 2 n \cos\Big( 2(N-n) \omega \Delta t\Big) \Big] \;,
\label{uno}
\end{equation}
vs.
\begin{equation}
|\xi (N,\omega \Delta t)|^2 = 2 \,(1- \cos 2N \omega \Delta t ) =
2 \,(1-\cos \omega (t_N-t_0)) \;,
\label{due} 
\end{equation}
where (\ref{csi}) and (\ref{etak}) have been used and an identical 
proportionality factor is understood. The unperturbed contribution $|\xi|^2$
simply oscillates between values $0,4$ with a period $(\pi/N)$. The function 
$|\eta|^2$, instead, is strongly oscillating for increasing $N$, developing
$2(N-1)$ local minima and a sharply-peaked absolute maximum at 
$\omega \Delta t=\pi$. Constructive interference is highest at the maximum, 
leading to a value $|\eta|^2_{\text{\small Max}}=16\,N^2$ that can 
be very large, while 
destructive interference strongly damps the function for $\omega \Delta t 
< \pi/2$, which is a zero for both $|\eta|^2$ and $|\xi|^2$ for any $N$. One 
can show that
\begin{equation}
|\eta(N, \omega \Delta t) |^2 \, \leq \, |\xi(N, \omega \Delta t) |^2 
\hspace{2cm}\text{on }[0, {\pi / 2} ] \text{ for any } N\;.
\label{diff} 
\end{equation}
Back to decoherence properties, condition
(\ref{diff}) means that, for a mode at frequency $\omega$, a finite region 
$\omega \Delta t \leq \pi/2$ exists, where the contribution to decoherence is 
{\sl smaller} in the presence of pulses. Since the ``correcting region'' is 
entered for small $\Delta t$ values, this effect takes place in the regime
of rapid flipping we expected. Since, moreover, smaller $\Delta t$ values 
require longer pulse sequences in order to evolve the system over the same 
interval, it is useful to consider an interesting limiting case.   

\newpage
\begin{center}
{\bf C. The limit of continuous flipping and suppression of decoherence}
\end{center}

Let us study an idealized situation represented by the following mathematical
limit:
\begin{equation}
\left\{ \begin{array}{c}
\Delta t \rightarrow 0 \;, \\
N \rightarrow \infty \;, \\
2N\Delta t = t_N-t_0 \;. 
\end{array} \right.
\label{limit}
\end{equation}
It is convenient to rewrite the decoherence function (\ref{gammap}) by 
exploiting (\ref{eta2}) to separate formally the unperturbed and the 
interference contributions. Thus
\begin{equation}
\Gamma_P(N,\Delta t)=\sum_k \, {|\xi_k(2N \Delta t)|^2 \over 2} \,
\coth\bigg( {\omega_k \over 2T}\bigg) \,\Big| 1-f_k(N,\Delta t) \Big|^2\;,
\label{gammap2}
\end{equation}
where
\begin{equation}
f_k(N,\Delta t)= 2\, {\xi_k(\Delta t) \over \xi_k(2N\Delta t)} 
\sum_{n=1}^N \,\mbox{e}^{2i (n-1)\omega_k \Delta t }\;.
\label{fk}
\end{equation}
In this way, $\Gamma_0(t_N-t_0)$ is recovered by putting $f_k=0$ for each 
mode, see (\ref{gamma0}). 
We evaluate the asymptotic limit of $f_k(N, \Delta t)$ as follows:
\begin{eqnarray}
\lim_{\Delta t \rightarrow 0} f_k(N, \Delta t) & = & 
{ \mbox{e}^{-i\omega_k t_0} \over 1-\mbox{e}^{i \omega_k(t_N-t_0)} }
\lim_{\Delta t \rightarrow 0} 
{ (1-\mbox{e}^{i \omega_k \Delta t}) \over \Delta t }
\sum_{n=1}^N \, 2 \Delta t \, \mbox{e}^{i \omega_k t_{n-1} }  \nonumber \\
& = & 
{ \mbox{e}^{-i\omega_k t_0} \over 1-\mbox{e}^{i \omega_k(t_N-t_0)} }
\lim_{\Delta t \rightarrow 0} 
{ (1-\mbox{e}^{i \omega_k \Delta t}) \over \Delta t }
\int_{t_0}^{t_N} ds\, \mbox{e}^{i \omega_k s}               \nonumber \\
& = &
\lim_{\Delta t \rightarrow 0} 
\bigg[ {\sin \omega_k \Delta t \over \omega_k \Delta t} + i \, 
       {1- \cos \omega_k \Delta t \over \omega_k \Delta t} \bigg] =\, 1 \;.
\label{limit2}
\end{eqnarray}
If this result holds for an arbitrary field mode, then the implications for 
the decoherence properties are transparent:
\begin{equation}
\lim_{\Delta t \rightarrow 0} \Gamma_P(N, \Delta t) =0\;, 
\label{suppress}
\end{equation}
i.e. in the limit of continuous flipping, decoherence is completely and 
exactly eliminated for any temperature and any spectral density function.  

Obviously, there is no hope that a continuous limit of this kind would be ever 
attained in practice. However, situation (\ref{limit}) should be approached 
if $\Delta t$ is made small compared to the 
fastest characteristic time present in the dynamics of the system. From 
the considerations of Sec. II, the environment correlation time $\tau_c$ 
certainly provides a lower bound since there is no spectral content of the 
environmental noise at frequency higher than $\omega_c$. Hence, we expect that
a sufficient condition in order to meet (\ref{limit}) is  
\begin{equation}
\omega_c \, \Delta t \lesssim 1 \;. 
\label{condition}
\end{equation}
The question on whether we can do better than this may be not completely 
obvious, since time scales different from $\tau_c$ are also involved in the 
decoherence process. In what follows, we try to understand this point both by
presenting a physical explanation for the decoherence suppression and by
analyzing some specific situations. 
   
\begin{center}
{\bf D. Physical interpretation}
\end{center}

People familiar with quantum Zeno effect \cite{zeno} may have found 
some similarities with the behavior we are discussing. In 
particular, the basic mathematical modification associated with pulses is 
a function $|\eta_k|^2$, which is O$(\omega_k^4 \Delta t^4)$ for short time 
intervals, 
compared to $|\xi_k|^2=\mbox{O}(\omega_k^2 \Delta t^2)$. Moreover,  
(\ref{limit}) is formally the same continuous limit involved in many quantum 
Zeno related proposals, notably the one by Cook \cite{cook}. In both cases, 
a preexisting dynamics - the qubit $+$ bath evolution here, a stimulated 
two-level transition in Cook's scheme - is modified through a pulsing 
procedure. Pulses respectively represent spin-flip interactions, short 
enough that the action of the bath is made negligible, or measurement pulses,
long enough that the coupling to the external environment (the measuring 
apparatus) is made appreciable. A dynamical inhibition phenomenon occurs when
pulses become sufficiently frequent. One can say that two opposite 
configurations are realized for decoherence: with continuous 
measurements, the interaction with the bath is always ``on'' and internal 
dynamics becomes frozen; in the limit of continuous flipping, it is the 
two-level controlled dynamics that dominates, and the interaction with the 
bath tends to be always ``off'', as indicated by (\ref{suppress}). 
However, the analogy stops from a more physical point of view.

A more interesting interpretation of the decoherence suppression 
(\ref{suppress}) can be obtained by connecting it to effects already observed
in magnetic resonance experiments, like spin-echoes, solid-echoes, or 
spin-flip narrowing \cite{slichter}. All of these are basically time-reversal
experiments. In order to capture the basic physical mechanism of our model,
let us go back to the elementaty spin-cycle (Sec. IIIA) and look more 
carefully at the evolution operator (\ref{timeop2}). This is made of two 
pieces: a free evolution during the first $\Delta t$, followed by an 
evolution governed by (\ref{transformed}) during the second interval 
$\Delta t$. It is this transformed operator that {\sl simulates} the
effect of a time-reversal. In fact, backward propagation during the second
part of the cycle would correspond to
\begin{equation}
\tilde{U}_{tot}(t_0+\Delta t, t_0)= \exp\bigg\{ - {\sigma_z \over 2} 
\sum_k \Big( b_k^\dagger \mbox{e}^{i \omega_k t_0} 
\xi_k(\Delta t) - b_k \mbox{e}^{-i \omega_k t_0} 
\xi_k^\ast(\Delta t) \Big) \bigg\} \;,
\label{back}
\end{equation}
By comparison with (\ref{transformed}), forward propagation in 
the presence of rf-kicks only differs for an additional term  
$\mbox{e}^{i \omega_k \Delta t}$ affecting each mode. This phase factor, 
due to the dynamics of the bath oscillators, is ultimately responsible
for the decoherence properties in the pulsed evolution of the system. If not 
for this phase difference between (\ref{transformed}) and (\ref{back}), we 
would have $\eta_k(\Delta t)=0$ and no decoherence. Instead, reversal is 
approximate since the bath restarts at time $t_0+\Delta t$ after the first 
pulse with a dephased initial condition (see also Appendix). However, 
if $\mbox{e}^{i \omega_k \Delta t} \approx 1$ for each mode, then the couple 
of kicks produces an exact time-reversal and, by iteration, we arrive at 
(\ref{suppress}). 
Equivalently, if the bath Hamiltonian can be considered as a constant, then 
the total Hamiltonian (\ref{hamiltonian}) acquires a minus sign and the 
system retraces the previous evolution. The validity of this
condition depends on the time scale we are considering. For a single mode of 
frequency $\omega$, the time needed to produce appreciable dephasing is 
$\tau  \approx \omega^{-1}$, so we expect decoherence correction 
for $\tau/\Delta t \gtrsim 1$. This is in agreement with both the 
interpretation of (\ref{diff}) and the semiclassical NMR argument 
\cite{slichter}, where motional effects are predicted for $\tau$ comparable 
to the mean time spent in a given spin configuration (here $\Delta t$). 
For the whole environment, the 
correlation time $\tau_c$ is the minimum time scale over which the 
dynamics is approximately unchanged, and the same reasoning leads to a 
physical explanation of the rapid flipping condition (\ref{condition}).

As a final remark, we point out that the mechanism accomplishing time-reversal 
in our model is purely {\sl macroscopic} in the sense that no reference is made to
the dynamical state of the system. This is different from the familiar case
of the Maxwell demon, that effectively reverses a dynamical evolution by 
operating over some microscopic variables (like velocities) at a given 
instant. Rather, the reversal is obtained by changing the sign of the system
Hamiltonian through the action of suitable external fields (control). A 
different kind of demon, the so-called Loschmidt demon, has been introduced
by some authors to account for this behavior \cite{pines}. In this 
terminology, the spin-flip procedure realizes in principle a Loschmidt demon
for a decohering qubit.   

\section{Analysis and examples}

In this section we try to give some semiquantitative picture of the 
decoherence mechanisms discussed so far. We focus our attention on a 
representative class of reservoirs, corresponding to so-called Ohmic 
environments. The appropriate spectral density is given by (\ref{density}) 
with $n=1$. The time dependence of the decoherence is 
summarized by the following expression:
\begin{equation}
\Gamma(t)=\alpha \, \int_0^\infty d\omega \,
\mbox{e}^{-\omega/\omega_c} \Big( 2\,\overline{n}(\omega,T) + 1 \Big) \,
{ 1- \cos \omega t \over \omega} \,
\Big| 1-f(\omega,N,\Delta t) \Big|^2\;,
\label{ohmic}
\end{equation}
where we assume $t_0=0$. Eq. (\ref{ohmic}) reproduces the unperturbed 
behavior of Sec. II when $f=0$, in which case $\Gamma(t) = \Gamma_0(t)$, 
Eq. (\ref{gamma0}). In the presence of $N$ spin-flip cycles, 
Eq. (\ref{ohmic}) is found from the continuous limit of $\Gamma_P(N,\Delta t)$
in (\ref{gammap}), with $f(\omega,N,\Delta t)$ given by (\ref{fk}). According
to Sect. III, in this case we are interested at decoherence after a complete 
pulse sequence, i.e. $t=t_N=2N \Delta t$.
For a fixed strength $\alpha$ of the system-reservoir coupling, the properties 
of the environment enter (\ref{ohmic}) with two parameters, 
$\omega_c \sim \tau_c^{-1}$, $T \sim \tau_{\beta}^{-1}$.

Let us first analyze decoherence in the absence of any correction, $f=0$. 
Qualitatively different behaviors arise depending on the relationship between
the cut-off frequency $\omega_c$ and the thermal frequency $\omega_{\beta}=T$.
Typically, two extreme situations are considered.
\par\noindent $(i)\;$ $\omega_c \ll \omega_{\beta}$: high-temperature limit or 
{\sl classical} environment:
\par\noindent 
Decoherence dynamics is relatively easy to describe in this regime 
since, due to the exponential dependence on the cut-off, $\omega_c$ is 
actually the only characteristic frequency accessible to the system. The
environment looks classical in the sense that its quantized structure cannot 
be appreciated compared to the thermal quantum $\omega_{\beta}$. Accordingly,
thermal fluctuations always dominate over vacuum ones and, after a short 
transient O$(\tau_{\beta})$ where decoherence is almost ineffective, dynamics
becomes very fast and coherence is lost completely 
after a time comparable to $\tau_c$. In general, one expects that in this 
limit equivalent results are obtained by assuming a heat bath of classical 
harmonic oscillators. A mixed quantum-classical derivation for a two-state
open system is given for instance in \cite{grigo}.
\par\noindent $(ii)$ $\omega_c \gg \omega_{\beta}$: low-temperature limit or
{\sl quantum} environment:
\par\noindent
In this limit a more complex interplay between thermal and vacuum effects 
arises. Thermal fluctuations are only effective for $t > \tau_{\beta}$ and, 
due to the exponential suppression of $\overline{n}(\omega,T)$, they are 
almost totally contributed by low frequency modes 
$\omega \lesssim \omega_{\beta}$. The effects of vacuum fluctuations dominate
on an intermediate region $\tau_c < t < \tau_{\beta}$, a nonvanishing  
contribution remaining, however, at longer time scales $t > \tau_{\beta}$.  
The frequency composition of the fluctuations is now less clear: while modes 
below the thermal threshold are still responsible for the long-time dynamics
$t > \tau_{\beta}$, frequencies in the range up to $\omega_c$ mostly 
contribute for $t < \tau_{\beta}$ but are also present beyond $\tau_{\beta}$.
As a consequence, even at thermal time scales $t > \tau_{\beta} \gg \tau_c$,
high frequency modes contribute appreciably to the decoherence process and 
characteristic times O$(\tau_c)$ are still relevant in the underlying 
dynamics.

Typical decoherence curves $\mbox{e}^{-\Gamma(t)}$ for the Ohmic environment 
are found by numerical integration of (\ref{ohmic}) and are shown in Fig. 1
for two choices corresponding to high- and low-temperature limit,
$\omega_c/T=10^{-2}$ and $\omega_c/T=10^2$ respectively. The partial 
contributions due to thermal and vacuum fluctuations are indicated separately 
where possible. 
In the low-temperature case, a quiet ($t<\tau_c$), a quantum 
($\tau_c < t < \tau_{\beta}$) and a thermal ($t > \tau_{\beta}$) regime are 
easily identified in the process, as indicated above and discussed in more 
detail by many authors \cite{unruh,palma,leggett}. In both configurations, 
the qubit coherence decays exponentially fast once the thermal regime is well
established,
\begin{equation}
\mbox{e}^{-\Gamma(t)} \approx \mbox{e}^{-t/t_{th}} \;,
\label{tth}
\end{equation}
for a suitable time constant $t_{th}$. In models where the decoherence rate
is constant in time, Eq. (\ref{tth}) is usually assumed as the definition of 
a typical decoherence time, $t_{dec}=t_{th}$. In our case, since the whole
behavior of $\Gamma(t)$ is required for a complete knowledge of the decoherence
dynamics, the definition of a characteristic time for loss of unitarity is 
less clear. Oversimplifying things, the situation can be 
summarized as follows: for both the high- and the low-temperature regimes, 
a characteristic time exists, indicating the departure of coherence from 
unity. This time is determined by the shortest between the two time scales 
$\tau_c, \tau_\beta$. Once this transient is over, the duration of the 
actual decoherence process is at least comparable to $\tau_c$ in the 
classical environment, and to $\tau_\beta$ in the quantum one. 
If this information 
is used as an estimate for a characteristic decoherence time, we are lead to 
$t_{dec}\approx \mbox{O}(\tau_c/\alpha)$ for the high-temperature limit and
$t_{dec}\approx \mbox{O}(\tau_\beta/\alpha)$ for the low-temperature limit
respectively.  For identical values of $\alpha$ and $\omega_c$, decoherence 
occurs extremely faster in the former case, as expected on intuitive grounds.

We now come back to examine how decoherence is improved in the presence of 
spin-flip cycles. With the interference contribution $f$ restored in 
(\ref{ohmic}), we have calculated numerically the decoherence values obtained 
when a fixed time interval $t$ is divided in an increasing number of cycles 
of duration $2 \Delta t$, i.e.
\begin{equation} 
\Delta t = {t \over 2N}\;, \hspace{2cm}N=1,2,\ldots,N_{\text{max}}\;.
\end{equation}
The behavior of $\mbox{e}^{-\Gamma(t)}$ as a function of the pulse frequency 
$1/\Delta t$ has been studied, and the procedure repeated for different
representative times. The results for the high- and low-temperature reservoirs
considered above are shown in Figs. 2, 3 respectively. The unperturbed values 
of decoherence at the appropriate times can be read from Fig. 1. We see that,
as predicted by (\ref{suppress}), decoherence-correction starts  
as soon as the region $\tau_c/\Delta t \gtrsim 1$ is entered. For times short 
enough that $\omega_c \,t < 2N$, this may by even accomplished with a 
single cycle. In general, once this condition is fulfilled, no further 
reduction of $\Delta t$ is demanded to evolve the system in a 
decoherence-free way. 
However, a warning also emerges from Fig. 3: if flipping is not frequent 
enough, not only does the correction effect disappear, but decoherence can be 
actually made worse compared to the one in the absence of pulses. 
The explanation of this behavior is rooted into the
interference mechanism that builds up the correction factor (\ref{eta}): in 
a sense, decoherence can be subtracted almost completely from the frequency
range $\omega \Delta t \lesssim 1$ only at the expense of enhancing the
decoherence contributions from modes outside that region. This intrinsic 
feature of the model is also relevant to understand why, despite of the 
differences existing between the high- and low-temperature decoherence 
properties, the same condition $\omega_c \Delta t \lesssim 1$ is required to 
prevent both thermal and vacuum noises. In fact, this condition comes quite
natural for a classical environment, but one might at first wonder why a 
weaker condition $\omega_{\beta} \Delta t \lesssim 1$ would not suffice for the
quantum case, even if the frequencies $\omega \lesssim  \omega_{\beta}$ 
contain the fraction that mostly contributes at times of the order of 
$t_{dec}$. The reason for this failure is the presence of 
vacuum fluctuations. By satisfying 
condition $\omega_{\beta} \Delta t \lesssim 1$, we do get rid of thermal 
dephasing, but we do not correct completely vacuum noise until 
$\omega_c \Delta t \lesssim 1$. Precisely, we are missing modes of 
intermediate frequency $\omega_{\beta} \lesssim \omega \lesssim \omega_c$
that, although of minor importance at long times, may introduce amplified 
decoherence contributions if not properly corrected.    

We conclude with a few comments on the relevance of our procedure for 
quantum information processing. While the idealized character of the 
model prevents us from a quantitative discussion of implementation criteria,
we can compare with the principles underlying current quantum error-correction 
proposals. Essentially, these are schemes to encode redundantly information
in such a way that it can be restored also when errors due to external 
sources have occurred. The syndrome-identification and the error-correction
stages may be regarded as a feedback configuration: suitable measurement 
protocols are required both in conventional schemes \cite{shor,steane,calderbank,laflamme,bennett,gottesman,cirac,divincenzo,schumacher} and in 
alternate techniques based on the quantum Zeno effect \cite{vaidman,luming},
while conditional logic is exploited in the coherence-preserving routines 
proposed in \cite{slotine}. In any case, error-correction methods have the 
effect of reducing the error rate per unit time and, in order to be effective, 
they must be repeated at time intervals $\Delta t$ shorter than
the typical decoherence time of the system, i.e. $t_{dec}/\Delta t
\gtrsim 1$. In comparison to quantum error correction schemes,
our procedure exhibits two fundamental 
differences: no ancillary bits are required to store the information; no 
measurements are performed. In principle, these could be advantageous 
features, since encoding would be more efficient and, by avoiding 
measurements, no slow-down of the computational speed would be introduced. In 
addition, rather than reducing the error rate, this method would suppress 
completely the error source provided the appropriate condition 
$\tau_c /\Delta t \gtrsim 1$ is fulfilled. 
From a more practical perspective, it is the accessibility of this rapid 
flipping limit, demanding fast and short pulses, 
$\tau_c /\Delta t \gtrsim 1$ and $\tau_P /\Delta t \ll 1$, that determines
the viability of the procedure itself. If such requirements can be satisfied,
our method might be valuable in configurations where $t_{dec}$ tends to be 
shorter than $\tau_c$ \cite{nota2} or, even in case $t_{dec}$ is longer 
compared to $\tau_c$, for systems where tipping the state is easier than 
exploiting conventional error-correction protocols. While the existence of
an interaction able to implement a NOT gate by inverting the state is 
demanded for any two-level system relevant to quantum computation, both the
present technological capabilities for applying $\pi$-pulses and the 
relevant environmental cut-off frequencies depend considerably on the 
specific physical system and the mechanism responsible for decoherence. Some
important time scales for various prospective qubits can be found in 
\cite{review}.
    
\section{Conclusions}

Our work demonstrates the possibility to modify the evolution of a quantum 
open system through the application of an external controllable interaction.
A prototype situation involving a two-level system coupled 
to a quantized reservoir in thermal equilibrium has been worked out in detail 
and dynamical suppression of quantum decoherence has been evidenced. From the 
perspective of quantum information, the analysis suggests a different 
direction compared to conventional quantum error-correction techniques, 
based on the idea of forcing the
system into a dynamics that disturbs the decoherence process. Our present 
study for a specific example brings up, among other issues, the question 
of whether similar decoherence correction mechanisms would be operating
under more general conditions, including either different open system 
dynamics, or different control configurations, or both. In particular, 
an interesting 
possibility could emerge from examining decoherence properties within a 
fully quantum mechanical description where the control degrees of freedom 
are explicitly included and the system is driven by a quantum controller as 
recently proposed in \cite{seth1}.

\acknowledgments
One of us (L. V.) is grateful to Carlo Presilla for enlightening 
discussions and a critical reading of the manuscript. 
This work was supported by ONR, by AFOSR, and by DARPA/ARO under the 
Quantum Information and Computation initiative (QUIC) and the NMR Quantum 
Computing initiative (NMRQC). 

\appendix
\section{Heisenberg representation}

Compared to the interaction picture, the Heisenberg representation has two 
advantages: first, it does not require preliminary transformations on the 
state vector; second, it gives to a certain extent a more intuitive 
description of the spin motion. In this Appendix, we outline the evolution of 
the qubit coherence, by restricting ourselves to the first elementary 
spin-flip cycle. In the Heisenberg picture, the relevant information is 
contained in 
\begin{equation}
\sigma_+(t) = {1 \over 2} \Big( \sigma_x(t) + i\, \sigma_y(t) \Big) \;, 
\label{a1}
\end{equation}
since, by averaging over the quantum state, $\langle \sigma_+(t) \rangle =
\rho_{01}(t)$. As in Sec. III, we evaluate the qubit dynamics under the 
separate action of the spin-bath Hamiltonian $H_{SB}$ and the radiofrequency
perturbation $H_{rf}$. The description of a $\pi$-pulse turns out to be 
extremely simple in the Heisenberg representation. Nothing happens to the 
bath operators $b_k,b^\dagger_k$ in the limit of instantaneous pulses, while 
spin dynamics is governed by the equations 
\begin{equation}
\dot{\sigma}_\alpha = -i\,\Big[\sigma_\alpha, \Big( 
H_S+H_{rf}(\omega_0,t) \Big) \Big]\;,   \hspace{2cm}\alpha=x,y,z\;,
\label{a2}
\end{equation}
with $H_{rf}(\omega_0,t)$ given in (\ref{rf}). By denoting with $t_P^{-\,(+)}$
the instants immediately before (after) a pulse respectively, a very simple
result is found:
\begin{equation}
\left\{ \begin{array}{lcl}
 \sigma_z(t^+_P) & = & -\, \sigma_z(t^-_P)\;, \\
 \sigma_+(t_P^+) & = & (\sigma_+(t_P^-))^\dagger\;. 
\end{array} \right.
\label{a3}
\end{equation}
The action on $\sigma_z$ corresponds, in particular, to the pictorial 
spin-flip effect operated by a $\pi$-pulse. Now denote as $G_{tot}(t_i,t_j)$
the operator evolving coherence from $t_i$ to $t_j$ in the absence of 
rf-pulses, i.e.
\begin{equation}
\sigma_+(t_j)= G_{tot}(t_i,t_j)\, \sigma_+(t_i) \;. 
\label{a4}
\end{equation}
Then, using relation (\ref{a3}) for $\sigma_+$ twice, we find the following 
representation for the coherence evolution during the first complete cycle:
\begin{equation}
\sigma_+(t_0 + 2 \Delta t) = G_{tot}(t_0,t_0+\Delta t) \, \sigma_+(t_0) \,
G_{tot}^\dagger (t_0+ \Delta t, t_0+ 2 \Delta t) \;, 
\label{a5}
\end{equation}
to be compared with 
\begin{equation}
\sigma_+(t_0 + 2 \Delta t) = G_{tot}(t_0+\Delta t, t_0+ 2\Delta t) 
\,G_{tot}(t_0, t_0+ \Delta t)\, \sigma_+(t_0)  
\label{a6}
\end{equation}
in the absence of pulses. Even before knowing the explicit form of $G_{tot}$, 
we see from (\ref{a5})-(\ref{a6}) that the presence of a time-reversed 
evolution during the second interval $\Delta t$ of the cycle is already 
enucleated at this stage.

In order to evaluate the propagator $G_{tot}$, the Heisenberg equations for 
the coupled spin $+$ bath motion have to be solved. From Hamiltonian 
(\ref{hamiltonian}) we get 
\begin{equation}
\left\{ \begin{array}{lcl}
\dot{b}_k & = & -i \omega_k b_k -i g_k \sigma_z \;, \\
\dot{b}_k^\dagger & = & +i \omega_k b^\dagger_k +i g_k^\ast \sigma_z \;, \\
\dot{\sigma}_+& = & i \omega_0 \sigma_+ + 2i \sum_k \, 
               (g_k b_k^\dagger + g_k^\ast b_k)\, \sigma_+ \;, \\
\dot{\sigma}_z & = & 0 \;.
\end{array} \right. 
\label{a7}
\end{equation}
Since instantaneous pulses introduce discontinuous changes in operators, the 
propagators $G_{tot}(t_0,t_0+\Delta t)$, 
$G_{tot}(t_0+\Delta t,t_0+2 \Delta t)$ have to be considered separately, by 
solving (\ref{a7}) with initial conditions at $t=t_0$, $t=t_P^+=t_0 + 
\Delta t$ respectively.
\par\noindent $(i)\;\; t_0 \mapsto t_0 +\Delta t$: \par\noindent
Since $\sigma_z(t)=\sigma_z(t_0)$, the equations for the bath variables are
completely solved by 
\begin{equation}
b_k(t) = \mbox{e}^{-i \omega_k(t-t_0)} b_k(t_0)- \sigma_z(t_0)
{g_k \over \omega_k} \Big( 1-\mbox{e}^{-i \omega_k (t-t_0)} \Big)\;, 
\label{a8}
\end{equation}
and $b_k^\dagger(t)=(b_k(t))^\dagger$. These solutions should be inserted in
the expression for $\sigma_+(t)$:
\begin{equation}
\sigma_+(t)=\mbox{T}\exp\bigg\{ i \int_{t_0}^t \, ds\,
\Big[ \omega_0 + 2(g_k b^\dagger_k(s) + g_k^\ast b_k(s) ) \Big] \bigg \} 
\,\sigma_+ (t_0) \;.  \label{a9}
\end{equation}
The time-ordered exponential can be evaluated exactly and the following result
is found for the first propagator:
\begin{eqnarray}
G_{tot}(t_0,t_0+\Delta t) & = &
\exp\bigg\{ i \omega_0 \Delta t + 4i {|g_k|^2 \over \omega_k} 
\Big(\Delta t - {\sin \omega_k \Delta t \over \omega_k} \Big)
(\openone -\sigma_z(t_0)) \bigg\} \nonumber \\
& \cdot & \exp\bigg\{ -  \sum_k  \Big( b_k^\dagger (t_0) \xi_k(\Delta t) - 
           b_k (t_0) \xi_k^\ast(\Delta t) \Big) \bigg\} \;, 
\label{a10}
\end{eqnarray}
where the same notation for $\xi_k(\Delta t)$ has been used, Eq. (\ref{csi}).
  
\par\noindent $(ii)\; t_0 + \Delta t \mapsto t_0 +2 \Delta t$: \par\noindent
By exploiting (\ref{a10}), we can immediately write down the expression for 
the propagator $G_{tot}^\dagger(t_0+\Delta t,t_0+2 \Delta t)$ in terms of 
the new initial condition at $t=t^+_P$:
\begin{eqnarray}
G_{tot}^\dagger(t_0+\Delta t,t_0+2\Delta t) & = &
\exp\bigg\{ - i \omega_0 \Delta t - 4i {|g_k|^2 \over \omega_k} 
\Big(\Delta t - {\sin \omega_k \Delta t \over \omega_k} \Big)
(\openone -\sigma_z(t_P^+)) \bigg\} \nonumber \\
& \cdot & \exp\bigg\{ + \sum_k  \Big( b_k^\dagger (t_P^+) \xi_k(\Delta t) - 
           b_k (t_P^+) \xi_k^\ast(\Delta t) \Big) \bigg\} \;. 
\label{a11}
\end{eqnarray}
Now everything can be evaluated with respect to the initial time of the cycle.
We exploit (\ref{a3}) for $\sigma_z(t^+_P)$ and, since $b_k(t_P^+)=b_k(t_P^-)=
b_k(t_0+ \Delta t)$, bath operators evolve as 
\begin{equation}
b_k(t_0+\Delta t) = \mbox{e}^{-i \omega_k \Delta t} b_k(t_0)- \sigma_z(t_0)
{g_k \over \omega_k} \Big( 1-\mbox{e}^{-i \omega_k \Delta t} \Big)\;, 
\label{a12}
\end{equation} 
and Hermitian conjugate. As we discussed in the text, the presence in 
(\ref{a12}) of an initial condition dephased by $\mbox{e}^{\pm i \omega_k 
\Delta t}$ for each environmental mode is the ultimate source of decoherence.
We arrive at the following expression for the cycle evolution:
\begin{eqnarray}
\sigma_+(t_0+2 \Delta t) & = &
\exp\bigg\{ i \varphi_1(\Delta t) (\openone -\sigma_z(t_0)) + \sum_k  
\Big( b_k^\dagger (t_0) \eta_k(\Delta t) - 
      b_k (t_0) \eta_k^\ast(\Delta t) \Big) 
+ i \varphi_2 (\Delta t) \bigg\} \nonumber \\
& \cdot & \sigma_+(t_0)
\exp\bigg\{ i \sigma_z(t_0) \varphi_2 (\Delta t) -
i \varphi_1(\Delta t) (\openone + \sigma_z(t_0)) \bigg\} \;, 
\label{a13}
\end{eqnarray}
where 
\begin{eqnarray}
\varphi_1(\Delta t) & = &
4 \sum_k {|g_k|^2 \over \omega_k} 
\Big(\Delta t - {\sin \omega_k \Delta t \over \omega_k} \Big) \;, 
\nonumber \\
\varphi_2(\Delta t) & = & 
8 \sum_k {|g_k|^2 \over \omega_k^2} \sin \omega_k \Delta t 
\, (1 - \cos \omega_k \Delta t) \;. 
\label{a14}
\end{eqnarray}
The final step is to calculate the coherence evolution as 
\begin{equation}
\rho_{01}(t_0+2\Delta t) = \langle \sigma_+(t_0+2\Delta t)\rangle =
\sum_{j=0,1} \langle j |\, \mbox{Tr}_B \{ \, (\rho_B \otimes \rho_S) \,
\sigma_+(t_0+2 \Delta t) \,\}\, | j \rangle \;. 
\label{a15}
\end{equation}
By inserting (\ref{a13})-(\ref{a14}), phase factors drop out and we find 
\begin{equation}
\rho_{01}(t_0+2\Delta t)= \rho_{01}(t_0) \, 
\mbox{e}^{- \Gamma_P (N=1, \Delta t)} \;,
\label{a16}
\end{equation}
in agreement with the result found from Eqs. (\ref{lab2})-(\ref{gammap}) in 
the Schr\"odinger representation. The procedure can be generalized to
arbitrary $N$ and the complete expression $\Gamma_P(N,\Delta t)$ is thereby 
recovered.

\newpage
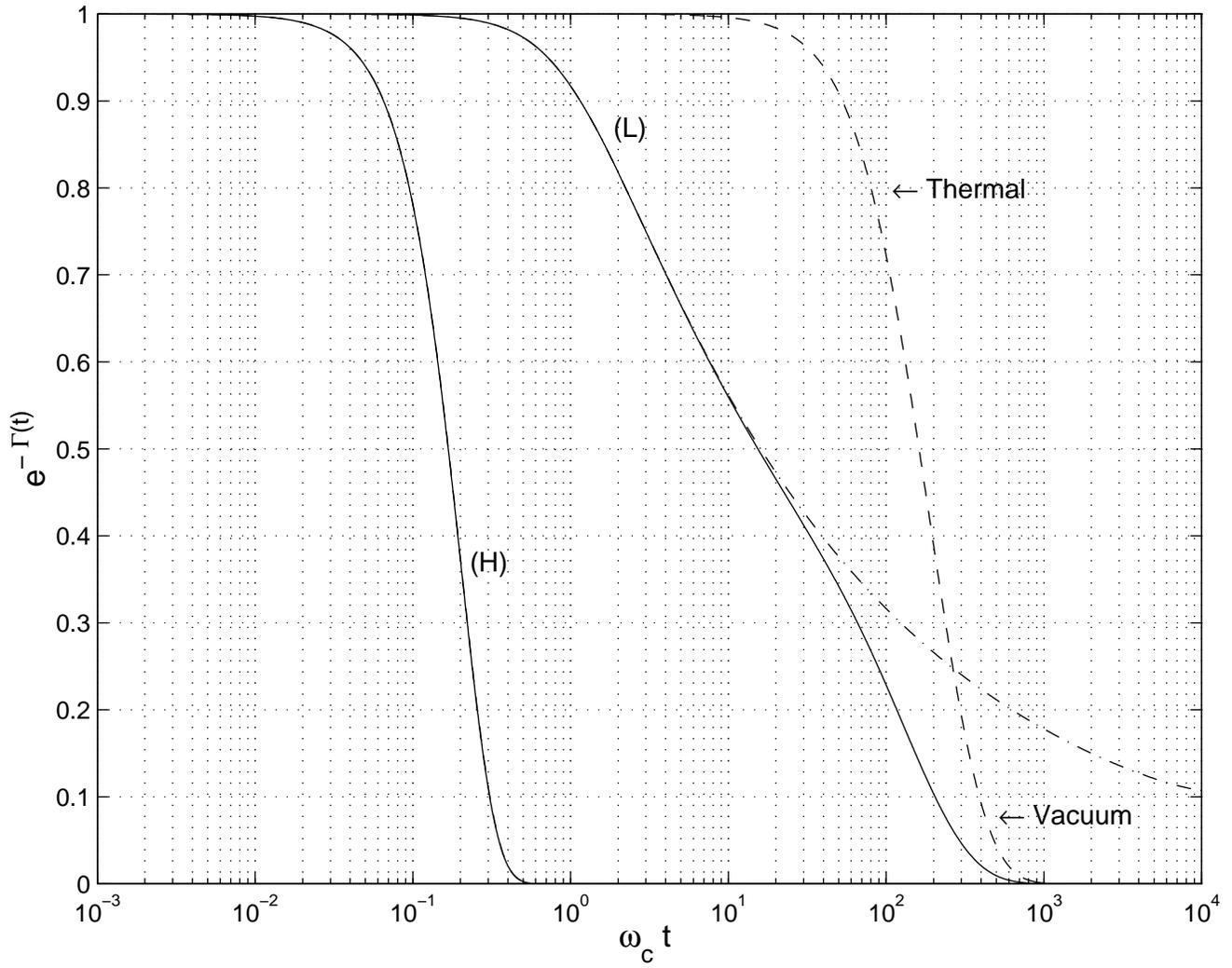
\begin{figure}
\caption{ Qubit decoherence as a function of time for an Ohmic environment, 
Eq. (52) ($f=0$). Time is in units of $T^{-1}$ and the values $\omega_c =100$,
$\alpha=0.25$ have been chosen. High- and low-temperature behaviors are 
shown, (H) $\omega_c/T=10^{-2}$ and (L) $\omega_c/T=10^2$ respectively. The 
contributions arising from the separate integration of thermal and vacuum 
fluctuations are displayed in the latter case, 
$\mbox{e}^{-\Gamma(t)}= \mbox{e}^{-\Gamma_{th}(t)} \cdot  
\mbox{e}^{-\Gamma_v(t)}$. }
\end{figure}    

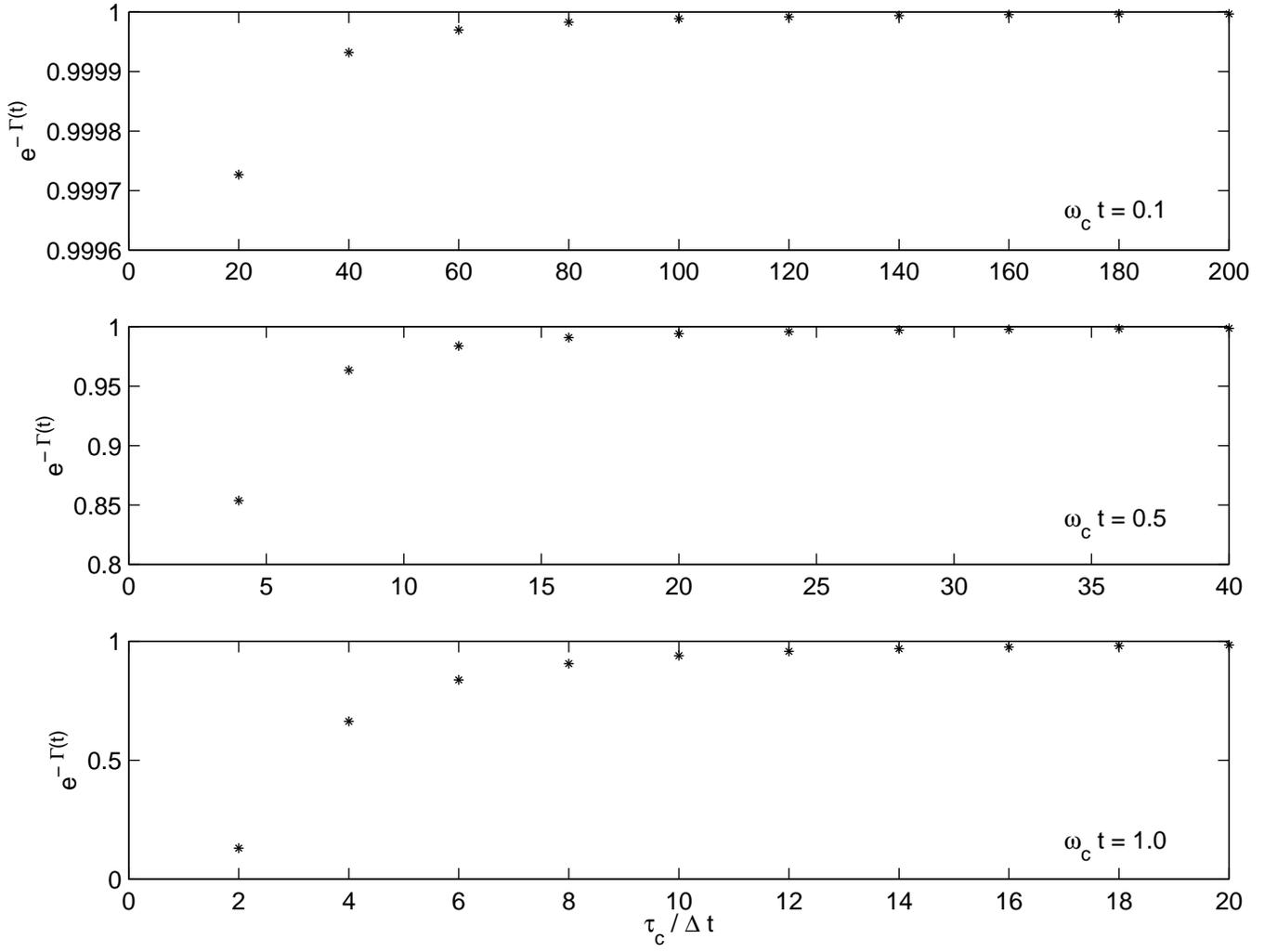
\begin{figure}
\caption{ Qubit decoherence in the presence of rf-pulses for the 
high-temperature configuration, $\omega_c/T=10^{-2}$. For a fixed time, each
point corresponds to a number $N$ of cycles, $N=1,\ldots,N_{\text{max}}=10$.
The results from Eq. (52) are plotted as a function of the normalized pulse
frequency $\tau_c /\Delta t$. The unperturbed values of decoherence are read 
from Fig. 1 (H). }
\end{figure}

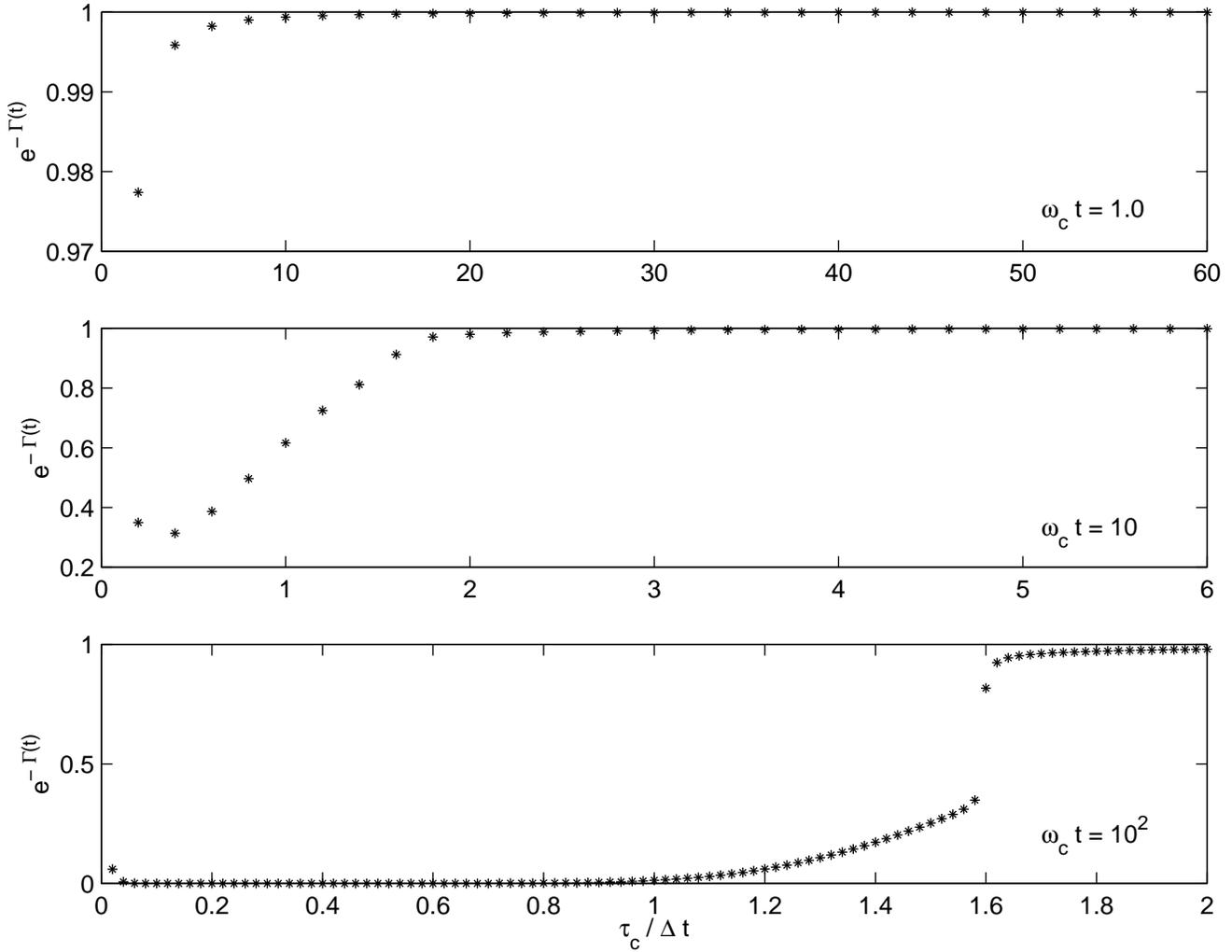
\begin{figure}
\caption{ Same as in Fig. 2 for the low-temperature configuration, 
$\omega_c/T=10^2$. The maximum number of spin-cycles is equal to 
$N_{\text{max}}=30$ in the simulations at $\omega_c \,t=1.0,10$, while 
$N_{\text{max}}=100$ at $\omega_c \, t=10^2$. The unperturbed values of
decoherence are read from Fig. 1 (L). }
\end{figure}   


\begin{references}
\bibitem{nota2} This may happen in the high-temperature and/or 
strong-coupling limit,  or for different classes of environment not 
considered here. See [3] for supra-Ohmic qubit decoherence.
\end{references}
\end{document}